\documentclass{kapproc} % Computer Modern font calls

\usepackage{amsmath}

\usepackage{procps} 

\usepackage[dvips]{graphicx}

\upperandlowercase

\normallatexbib %will produce bibliography entries as shown in the
\newcommand{\Jh}[0]{\hat{J}}
\newcommand{\jh}[0]{\hat{j}}
\newcommand{\sh}[0]{\hat{S}}
\newcommand{\ks}[0]{\kappa^2}

\renewcommand{\S}[1]{\hat{S}_{#1}}
\newcommand{\X}[1]{\hat{X}_{#1}}
\renewcommand{\P}[1]{\hat{P}_{#1}}
\newcommand{\J}[1]{{\hat{J}}_{#1}}

\newcommand{\commutator}[2]{\left[#1,#2\right]}
\newcommand{\mean}[1]{\left<#1\right>}
\newcommand{\var}[0]{\mathrm{Var}}
\newcommand{\vac}[1]{{\hat{V}}_{#1}}

\begin{document}

\articletitle{Distant Entanglement of Macroscopic Gas Samples }
\author{J. Sherson$^1$, B. Julsgaard$^2$, and E. S. Polzik$^2$} 
\affil{Danish National Research Foundation, Center for Quantum Optics
  - QUANTOP, (1) Department of Physics and Astronomy, University of
  Aarhus, Ny Munkegade bygning 520, 8000 Aarhus, Denmark. (2) Niels
  Bohr Institute, Blegdamsvej 17, 2100 Copenhagen, Denmark.}
\email{}

\begin{abstract}
  One of the main ingredients in most quantum information protocols is
  a reliable source of two entangled systems. Such systems have been
  generated experimentally several years ago for light
  \cite{aspect:82,shih:88,ou:92,kwiat:95,schori:02a} but has only in
  the past few years been demonstrated for atomic systems
  \cite{hagley:97,sackett,nature,roos:04}. None of these approaches
  however involve two atomic systems situated in separate
  environments. This is necessary for the creation of entanglement
  over arbitrary distances which is required for many quantum
  information protocols such as atomic teleportation
  \cite{bennett,kuzmich}. We present an experimental realization of
  such distant entanglement based on an adaptation of the entanglement
  of macroscopic gas samples containing about $10^{11}$ cesium atoms
  shown in \cite{nature}. The entanglement is generated via the
  off-resonant Kerr interaction between the atomic samples and a pulse
  of light.  The achieved entanglement distance is 0.35m but can be
  scaled arbitrarily. The feasibility of an implementation of various
  quantum information protocols using macroscopic samples of atoms has
  therefore been greatly increased.  We also present a theoretical
  modeling in terms of canonical position and momentum operators
  $\X{}$ and $\P{}$ describing the entanglement generation and
  verification in presence of decoherence mechanisms.
\end{abstract}

\section{Introduction}
Ever since Einstein, Podolsky, and Rosen in their seminal paper from
1935 \cite{EPR} introduced the possibility of entangling two quantum
system, entanglement has been viewed as one the most curious and
spectacular phenomena in quantum mechanics.  In the past few years
the role of entanglement in quantum mechanics has shifted dramatically
from being a fundamental test of the foundation of the entire quantum
mechanical theory to being a technical resource in the rapidly
developing field of quantum information. Thus, the hunt is on for
reliable sources of entanglement. These have been available for
discrete as well as continuous states of the electromagnetic field.
However, entangled states of material particles have presented a
greater experimental challenge.

Macroscopic samples of atoms as a resource of entanglement have
attracted a lot of attention in recent years because of their relatively
simple experimental realization (works at room temperature) and
robustness to single particle decoherence. In \cite{nature}
entanglement of this kind was accomplished. However, the two samples
were located only 1cm apart in the same shielded environment. This
meant that this implementation did not incorporate the important
feature in e.g.~the teleportation protocol, that the distance of
teleportation given by the separation of the entangled systems could
be arbitrary. We have created two separate environments each
containing a gas sample of cesium atoms at room temperature. As we
will see, we have successfully created entangled states between two
such systems being 0.35m apart. In the present paper we will also
focus on a better method to verify the generation of entangled states
as compared to the experiment in \cite{nature}.

\section{Light Atom Interaction}
In this section we introduce the physical systems involved in the
experiment, i.e.~we introduce the atomic spin samples and the
polarization state of laser pulses interacting with each other. Based
on the equations of motion we will in Sec.~\ref{sec:ent_create_model}
explain how the interaction can be utilized for entanglement
generation.

\subsection{Atomic System}
Our atomic system is composed of two separate samples of spin
polarized cesium vapour placed in paraffin coated glass cells at room
temperature. Cesium has a hyperfine split ground state with total
angular momentum $F=3$ and $F=4$, the latter being our atomic quantum
system of interest.  Having a macroscopic ensemble of atoms (around
$10^{11}$) we will define the theoretically discrete but effectively
continuous collective spin variables $\Jh_k=\sum_i \jh_k^i$, where
$k=x,y,z$ and $i$ denotes the individual atom. These will retain
regular angular momentum commutation relations,
$[\Jh_y,\Jh_z]=i\Jh_x$.  The Heisenberg uncertainty relation will then
lead to $\textrm{Var}(\Jh_y)\textrm{Var}(\Jh_z)\geq
\frac{1}{4}|\left<\Jh_x\right>|^2$. We will be interested in the state
in which practically all atoms are in the $F=4$, $m_F=4$ state with
$x$ as quantization axis. All the relevant interactions only
involve minute changes in the macroscopic spin orientation. $\Jh_x$ can
then be regarded as a constant classical number. If all atoms are
independent this will give rise to a minimum uncertainty state called
the coherent spin state (CSS) with:
\begin{equation}
\label{eq:varcss}
\qquad \textrm{Var}(\Jh_y)=\textrm{Var}(\Jh_z)=J_x/2=2N_{\mathrm{atoms}}.
\qquad\qquad \textrm{(CSS)}
\end{equation}
We therefore see that the variance, often referred to as the projection
noise, of the CSS will grow proportionally to the number of atoms.
This scaling is manifestly quantum and will be important for the
verification of the entanglement.

\subsection{Light System}
In complete analogy to the atomic collective spin variables, the
polarization state of a pulse of light can be described by a vector,
the so-called Stokes vector. For light propagating along the $z$-axis
we define
\begin{equation}
    \S{x}(t) = \frac{1}{2}\left(\phi_x - \phi_y\right),\;
    \S{y}(t) = \frac{1}{2}\left(\phi_{45} - \phi_{135}\right),\;
    \S{z}(t) = \frac{1}{2}\left(\phi_+ - \phi_-\right),
\end{equation}
where $\phi_x$, $\phi_y$ are the photon fluxes of $x$ and
$y$-polarized photons, $\phi_{45}$, $\phi_{135}$ are photon fluxes
measured in a basis rotated $45^{\circ}$ with respect to the
$x,y$-axes, and $\phi_+$, $\phi_-$ refer to photons with $\sigma_+$
and $\sigma_-$-polarization. In our experiments the light will with
very good approximation be linearly polarized along the $x$-axis.
Then $\S{x}(t)$ can be described by a classical $c$-number. The
$\S{y}(t)$ and $\S{z}(t)$ operators will contain the interesting
quantum variables. The Stokes operators defined above have dimension
$\textrm{time}^{-1}$. This is convenient for describing light/matter
interactions as we will see below. But for entanglement generation or
quantum information protocols in general it is more convenient to
consider entire pulses which are time integrated versions of the
above. It can be shown that
\begin{equation}
\commutator{\int_0^T \S{y}(t)dt}{\int_0^T\S{z}(t)dt} = i\int_0^T\S{x}(t)dt
  = n_{\mathrm{ph}}/2,
\end{equation}
where the last equality holds in our case for strong linear
polarization along the $x$-axis. From this commutator we derive the
variance of the minimum uncertainty state called the coherent state:
\begin{equation}
\label{eq:shot_noise}
  \var\left(\int_0^T \S{y}(t)dt\right) = \var\left(\int_0^T \S{z}(t)dt\right)
    = n_{\mathrm{ph}}/4. \quad \textrm{(shot noise)}
\end{equation}
Note again the characteristic linear quantum scaling with the number
of particles $n_{\mathrm{ph}}$. We often refer to the variance of the
coherent light state as the shot noise level. 

For completeness we will note that for coherent light states it can be
shown \cite{julsgaard_qic} that
\begin{equation}
\label{eq:correlation_Syz}
  \mean{\S{y}(t)\S{y}(t')} = \mean{\S{z}(t)\S{z}(t')} 
    = \frac{S_x}{2}\delta(t-t').
\end{equation}
The usefulness of this will be shown below. In our experiment we will
use $\sh_y$ detections to create entanglement between atomic samples.
If initially $x$-polarized light is rotated a small angle $\theta$
around the axis of propagation ($z$) we get
$\left<\hat{S}_y\right>=2S_x\theta$. This is why an $\sh_y$ detection
is sometimes referred to as a polarization rotation measurement.
\subsection{Interaction}
We couple our light and atomic system by tuning a laser beam
off-resonantly to the $6S_{1/2} \rightarrow 6P_{3/2}$ dipole
transition in cesium. This leads to the following equations of
interaction:
\begin{eqnarray}
\label{eq:syout}
  \hat{S}_y^{\mathrm{out}}(t) &=& \hat{S}_y^{\mathrm{in}}(t)
+a S_x \hat{J}_z(t), \\
\label{eq:szout}
  \hat{S}_z^{\mathrm{out}}(t) &=& \hat{S}_z^{\mathrm{in}}(t), \\
\label{eq:jyout}
  \frac{\partial}{\partial t}\hat{J}_y(t) &=& 
a J_x \hat{S}_z^{\mathrm{in}}(t), \\
\label{eq:jzout}
  \frac{\partial}{\partial t}\hat{J}_z(t) &=& 0,
\end{eqnarray}
where $a = - \frac{\gamma \lambda^2}{8\pi A \Delta}$. $A$ is the beam
cross section, $\Delta$ is the detuning (red positive), $\lambda$ is
the optical wavelength, and $\gamma$ is the natural linewidth of the
excited state $6P_{3/2}$. In and out refer to light before and after
passing the atomic sample, respectively. The above equations have been
derived carefully in \cite{julsgaard_qic} but we will give a short physical
explanation here.

First of all, the interaction is refractive in nature (the absorption
of the off-resonant light is negligible). It is convenient to consider
the incoming linearly polarized light in the $\sigma_+$ and $\sigma_-$
basis. The phase shift of a $\sigma_+$ and a $\sigma_-$ photon
propagating through atoms will be different if there is a spin
component $\J{z}$ along the propagation direction. For instance
(quantized along $z$) an atom in the $m_F = F$ magnetic sub-level
couples strongly to $\sigma_+$ photons and weakly to $\sigma_-$
photons. For the $m_F = -F$ sub-level the situation is reversed. The
differential phase shift of $\sigma_+$, $\sigma_-$ photons turns out
to depend linearly on $\J{z}$ which leads to a polarization rotation
of the incoming linearly polarized light proportionally to $\J{z}$
(also known as Faraday rotation).  Eq.~(\ref{eq:syout}) is a first
order approximation of this effect.

The different coupling strengths for different sub-levels also lead
to a Stark shift of atomic levels depending on the sub-level quantum
number $m_F$ and the incoming light polarization. Integrated over
time, the Stark shifts lead to different phase changes of the magnetic
sub-levels which changes the spin state. If for instance there are
more $\sigma_+$ than $\sigma_-$ photons, the $m_F = F$ sub-state will
be affected more than the $m_F = -F$ state. The amount of $\sigma_+$
and $\sigma_-$ photons is measured by the $\S{z}$ operator. It turns
out that the spin state evolution can be described as a rotation of
the spin around the $z$-axis by an amount proportional to $\S{z}$.
This is given to first order in Eq.~(\ref{eq:jyout}) (the rotation is
so small that $J_x$ is unaffected).

In the interaction process the $\sigma_+$ and $\sigma_-$ photons
experience phase shifts but are not absorbed. In the off-resonant
limit the flux of $\sigma_+$ and $\sigma_-$ photons are individually
conserved leading to Eq.~(\ref{eq:szout}). By conservation of angular
momentum along the $z$-direction this leads to the constancy of
$\J{z}$ expressed by Eq.~(\ref{eq:jzout}).

We see from Eqs.~(\ref{eq:syout}) and (\ref{eq:jzout}) that in the
case of a large interaction strength (i.e.~if $aS_x\J{z}$ dominates
$\S{y}^{\mathrm{in}}$) a measurement on $\sh_y^{\mathrm{out}}$ amounts
to a measurement of $\Jh_z$ without destroying the state of $\Jh_z$.
This is termed a Quantum Non Demolition (QND) measurement of $\Jh_z$.
Using off-resonant light for QND measurements of spins has also been
discussed in \cite{kuzmich:98,takahashi:99}. We note that
Eq.~(\ref{eq:jyout}) implies that a part of the state of light is also
mapped onto the atoms. This opens up the possibility of using this
sort of system for quantum memory. One step in this direction is
discussed in \cite{julsgaard_qic,schori:02b}.

\subsection{Adding a Magnetic Field}
\label{sec:addbfield}
In the experiment a constant and homogeneous magnetic field is added in
the $x$-direction. We discuss the experimental reason for this below.
For our modeling the magnetic field adds a term $H_B=\Omega J_x$ to
the Hamiltonian. This makes the transverse spin components precess at
the Larmor frequency $\Omega$ depending on the strength of the field.
Introducing the rotating frame coordinates:
\begin{equation}
\left(
\begin{array}{c}
  \Jh_{y}' \\
  \Jh_{z}' \\
\end{array}
\right)
=
\left(
\begin{array}{cc}  \cos(\Omega t) & \sin(\Omega t) \\
  -\sin(\Omega t) & \cos(\Omega t) \\
\end{array}
\right)
\left(
\begin{array}{c}
  \Jh_{y} \\
  \Jh_{z} \\
\end{array}
\right)
\end{equation}
we can easily show that Eqs.~(\ref{eq:syout})-(\ref{eq:jzout}) will
transform into:
\begin{align}
\label{eq:Syout_rotating}
  \S{y}^{\mathrm{out}}(t) &= \S{y}^{\mathrm{in}}(t) + 
    a S_x\left(\J{y}'(t)\sin(\Omega t) + \J{z}'(t)\cos(\Omega t)\right), \\
\label{eq:Szout_rotating}
  \S{z}^{\mathrm{out}}(t) &= \S{z}^{\mathrm{in}}(t), \\
\label{eq:Jydot_rotating}
  \frac{\partial}{\partial t} \J{y}'(t)
    &= a J_x \S{z}^{\mathrm{in}}(t)\cos(\Omega t), \\
\label{eq:Jzdot_rotating}
  \frac{\partial}{\partial t} \J{z}'(t) &= a J_x
    \S{z}^{\mathrm{in}}(t)\sin(\Omega t).
\end{align}
Thus, the atomic imprint on the light is encoded in the
$\Omega$-sideband instead of at the carrier frequency. The advantage
of this added feature is threefold. The first and perhaps most
important advantage is that lasers are generally a lot more quiet at
high sideband frequencies compared to the carrier. A measurement
without a magnetic field will be a DC measurement and the technical
noise would dominate the subtle quantum signal. Secondly, the
$B$-field introduces a Larmor splitting of the magnetic sublevels of
the hyperfine ground state multiplet, thus lifting the degeneracy.
This will introduce an energy barrier strongly suppressing spin
flipping collision. The lifetime of the atomic spin state is
consequently greatly increased. The last advantage is that as long as
the measurement time is longer than $1/\Omega$
Eq.~(\ref{eq:Syout_rotating}) enables us to access both $J_{y}'$ and
$J_{z}'$ at the same time. We are of course not allowed to perform
non-destructive measurements on these two operators simultaneously since
they are non-commuting. This is also reflected by the fact that
neither $\J{y}$ nor $\J{z}$ are constant in
Eqs.~(\ref{eq:Jydot_rotating}) and~(\ref{eq:Jzdot_rotating}). In
section~\ref{sec:ent_create_model} we shall consider two atomic
samples and the third advantage becomes evident.

\section{Entanglement Creation and Modeling}
\label{sec:ent_create_model}
In this section we define what is meant by entangled states of atomic
samples and we adapt the equations of motion from previous sections
for this purpose. We will derive a simple model for the entanglement
creation and describe how to verify that the states created really are
entangled. In \cite{myart} a much more elaborate description of
entanglement creation is given.

\subsection{Entanglement Criterion}
\label{sec:entanglement-criterion}
Let us here state the criterion to fulfil in order to prove the
generation of entangled states.  Since entanglement is the nonlocal
interconnection of two systems we need to have two atomic samples A
and B.  Entanglement is usually defined in terms of density matrices
so that A and B are entangled if they are connected in such a way that
it is impossible to write the total density matrix as a product,
$\rho_{tot}=\sum p_i \rho_{Ai}\rho_{Bi}$. For our continuous variable
system we have the experimentally practical criterion derived from the
above definition in \cite{duanentcrit}:
\begin{equation}
\label{eq:entcritduan}
\textrm{Var}(\J{y1}'+\J{y2}')+\textrm{Var}(\J{z1}'+\J{z2}')<2J_x
\end{equation}
where we have assumed both samples to be macroscopically oriented with
same magnitude $J_x$. The two samples are indexed by 1 and 2.
Comparing to Eq.~(\ref{eq:varcss}) we get an equality for two
independent atomic samples in the CSS. To have entanglement we thus
need to know the sums of the spin components along the $y$- and the
$z$-directions better than we could ever know each of the spin
projections by itself. It is now interesting to examine the commutator
between the sums,
$\commutator{\J{y1}'+\J{y2}'}{\J{z1}'+\J{z2}'}=i(J_{x1}+J_{x2})$.  A
non-zero commutator means that increasing our knowledge of one
component will automatically decrease our knowledge of the other, thus
making our attempts to break the inequality of
Eq.~(\ref{eq:entcritduan}) futile.  Now comes the trick that makes
entanglement generation possible in our experiment. Assume $J_{x1}$
and $J_{x2}$ to be equal in magnitude but opposite in direction. The
commutator will then become zero and we can at least theoretically
measure both components with arbitrary precision, thereby satisfying
the entanglement criterion of Eq.~(\ref{eq:entcritduan}).

\subsection{Two Oppositely Oriented Spins}
Inspired by the above we will from now on assume $J_{x1} = -J_{x2}
\equiv J_x$. We will re-express the equations of
motion~(\ref{eq:Syout_rotating})-(\ref{eq:Jzdot_rotating}) for two
samples in a way which is much more convenient for the understanding
of our entanglement creation and verification procedure. We introduce
position and momentum like operators $\X{}$ and $\P{}$ to describe
pulses of light and the atomic systems. This is a more abstract but
hopefully also more well known and intuitive way to express the
interactions creating the entangled states.

For two atomic samples we write equations of motion:
\begin{align}
\label{eq:Syout_rotating_two}
  \S{y}^{\mathrm{out}}(t) &= \S{y}^{\mathrm{in}}(t) + 
    a S_x\left([\J{y1}'(t)+\J{y2}'(t)]\sin(\Omega t)\right. \\ \notag
      &\left.\qquad\qquad\qquad\; + [\J{z1}'(t)+\J{z2}'(t)]\cos(\Omega t)\right), \\
\label{eq:Jy1y2dot_rotating}
  \frac{\partial}{\partial t} (\J{y1}'(t)+\J{y2}'(t))
    &= a (J_{x1}+J_{x2})\S{z}^{\mathrm{in}}(t)\cos(\Omega t) = 0, \\
\label{eq:Jz1z2dot_rotating}
  \frac{\partial}{\partial t} (\J{z1}'(t)+\J{z2}'(t)) 
    &= a (J_{x1}+J_{x2})\S{z}^{\mathrm{in}}(t)\sin(\Omega t) = 0.
\end{align}
The fact that the sums $\J{y1}'(t)+\J{y2}'(t)$ and
$\J{z1}'(t)+\J{z2}'(t)$ have zero time derivative relies on the
assumption of opposite spins of equal magnitude. The constancy of
these terms together with Eq.~(\ref{eq:Syout_rotating_two}) allows us
to perform QND measurements on the two sums. We note that each of the
sums can be accessed by considering the two operators
\begin{align}
  \int_0^T \S{y}^{\mathrm{out}}\cos(\Omega t)dt &= 
    \int_0^T \S{y}^{\mathrm{in}}\cos(\Omega t)dt 
    + \frac{a S_x}{2}(\J{z1}'(t)+\J{z2}'(t)), \\
  \int_0^T \S{y}^{\mathrm{out}}\sin(\Omega t)dt &= 
    \int_0^T \S{y}^{\mathrm{in}}\sin(\Omega t)dt 
    + \frac{a S_x}{2}(\J{y1}'(t)+\J{y2}'(t)).
\end{align}
We have used the fact that $\int_0^T\cos^2(\Omega t)dt \approx
\int_0^T\sin^2(\Omega t)dt \approx 1/2$ and that $\int_0^T\cos(\Omega
t)\sin(\Omega t)dt \approx 0$. Each of the operators on the left hand
side can be measured simultaneously by making a $\S{y}$-measurement
and multiplying the photocurrent by $\cos(\Omega t)$ or $\sin(\Omega
t)$ followed by integration over the duration $T$. The possibility to
gain information about $\J{y1}'(t)+\J{y2}'(t)$ and
$\J{z1}'(t)+\J{z2}'(t)$ enables us to break the
inequality~(\ref{eq:entcritduan}). At the same time we must loose
information about some other physical variable. This is indeed true,
the conjugate variables to these sums are $\J{z2}'(t)-\J{z1}'(t)$ and
$\J{y1}'(t)-\J{y2}'(t)$, respectively. These have the time evolution
\begin{align}
  \frac{\partial}{\partial t} (\J{y1}'(t)-\J{y2}'(t))
    &= 2a J_x\S{z}^{\mathrm{in}}(t)\cos(\Omega t), \\
  \frac{\partial}{\partial t} (\J{z1}'(t)-\J{z2}'(t)) 
    &= 2a J_x\S{z}^{\mathrm{in}}(t)\sin(\Omega t).  
\end{align}
We see how noise from the input $\S{z}$-variable is piling up in the
difference components while we are allowed to learn about the sum
components via $\S{y}$ measurements. The above equations clearly
describe the physical ingredients in play but the notation is
cumbersome. Therefore we define new operators. For the atomic system
we take
\begin{subequations}
\begin{align}
  \X{A1} &= \frac{\J{y1}'-\J{y2}'}{\sqrt{2J_x}},  \\
  \P{A1} &= \frac{\J{z1}'+\J{z2}'}{\sqrt{2J_x}},  \\
  \X{A2} &= -\frac{\J{z1}'-\J{z2}'}{\sqrt{2J_x}}, \\
  \P{A2} &= \frac{\J{y1}'+\J{y2}'}{\sqrt{2J_x}}.  
\end{align}
\end{subequations}
New light operators will be
\begin{subequations}
\begin{align}
\label{eq:def_XL1}
  \X{L1} &= \sqrt{\frac{2}{S_x T}}\int_0^T\S{y}(t)\cos(\Omega t)dt, \\
  \P{L1} &= \sqrt{\frac{2}{S_x T}}\int_0^T\S{z}(t)\cos(\Omega t)dt, \\
  \X{L2} &= \sqrt{\frac{2}{S_x T}}\int_0^T\S{y}(t)\sin(\Omega t)dt, \\
  \P{L2} &= \sqrt{\frac{2}{S_x T}}\int_0^T\S{z}(t)\sin(\Omega t)dt.
\end{align}
\end{subequations}
Each pair of $\X{},\P{}$ operators satisfy the usual commutation
relation, e.g.~we have $\commutator{\X{L1}}{\P{L1}} = i$. All previous
equations now translate into
\begin{subequations}
\begin{align}
\label{interact_light}
  \X{Li}^{\mathrm{out}} &= \X{Li}^{\mathrm{in}} + \kappa\P{Ai}^{\mathrm{in}},\\
  \P{Li}^{\mathrm{out}} &= \P{Li}^{\mathrm{in}},\\
\label{interact_atom}
  \X{Ai}^{\mathrm{out}} &= \X{Ai}^{\mathrm{in}} + \kappa\P{Li}^{\mathrm{in}},\\
  \P{Ai}^{\mathrm{out}} &= \P{Ai}^{\mathrm{in}},
\end{align}
\end{subequations}
where we remember $i=1,2$ refer to the definitions above and not the
two samples. The parameter describing the strength of
light/matter-interactions is given by $\kappa = a\sqrt{J_x S_x T}$.
The limit to strong coupling is around $\kappa \approx 1$. Note, we
have two decoupled sets of interacting light and atomic operators.

In the transition from Stokes operators to canonical variables in
Eqs.~(\ref{eq:def_XL1}-d) the result~(\ref{eq:correlation_Syz}) is a
convenient tool for calculating variances. If for instance the input
light state is the coherent vacuum state we have
\begin{equation}
\begin{split}
  \var(\X{L1}) &= \mean{\X{L1}^2} \\
    &= \frac{2}{S_x T}\int_0^T\int_0^T 
    \mean{\S{y}^{\mathrm{in}}(t)\S{y}^{\mathrm{in}}(t')}
    \cos(\Omega t)\cos(\Omega t')dtdt' = \frac{1}{2}
\end{split}
\end{equation}
which is as expected. Likewise, if the two atomic samples are each in
the coherent state we will derive e.g.~$\var(\X{A1}) = 1/2$. Coherent
states of atomic or light systems as defined above correspond to what
is known as coherent states of the $\X{},\P{}$-operators.

The entanglement criterion~(\ref{eq:entcritduan}) written in
$\X{},\P{}$-language is
\begin{equation}
\label{eq:ent_critetrion_xp}
  \var(\P{A1}) + \var(\P{A2}) < 1.
\end{equation}
We see that entanglement of the two atomic samples can be considered
as so-called two mode squeezing. The uncertainty in the uncoupled pair
of operators $\P{A1}$ and $\P{A2}$ is reduced on the expense of the
increased noise in the operators $\X{A1}$ and $\X{A2}$.

\subsection{Entanglement Generation and Verification}
\label{sec:ent_gen_ver}
Now we turn to the actual understanding of entanglement generation and
verification. Experimentally we perform the following steps (more
details will be given in Sec.~\ref{sec:experimental-setup}). First the
atoms are prepared in the oppositely oriented coherent states
corresponding to creating the vacuum states of the two modes
$(\X{A1},\P{A1})$ and $(\X{A2},\P{A2})$. Next a pulse of light called
the \emph{entangling pulse} is sent through atoms and we measure the
two operators $\X{L1}^{\mathrm{out}}$ and $\X{L2}^{\mathrm{out}}$ with
outcomes $A_1$ and $B_1$, respectively.  These results bear
information about the atomic operators $\P{A1}$ and $\P{A2}$ and hence
we reduce variances $\var(\P{A1})$ and $\var(\P{A2})$. To prove we
have an entangled state we must confirm that the variances of $\P{A1}$
and $\P{A2}$ fulfil the criterion~(\ref{eq:ent_critetrion_xp}). That
is we need to know the mean values of $\P{A1}$ and $\P{A2}$ with a
total precision better than unity. For this demonstration we send a
second \emph{verifying pulse} through the atomic samples again
measuring $\X{L1}^{\mathrm{out}}$ and $\X{L2}^{\mathrm{out}}$ with
outcomes $A_2$ and $B_2$. Now it is a matter of comparing $A_1$ with
$A_2$ and $B_1$ with $B_2$. If the results are sufficiently close the
state created by the first pulse was entangled.

Now let us be more quantitative. The
interaction~(\ref{interact_light}) mapping the atomic operators
$\P{Ai}$ out on light is very useful for a strong $\kappa$ and useless
if $\kappa \ll 1$. We will describe in detail the role of $\kappa$ for
all values. To this end we first describe the natural way to determine
$\kappa$ experimentally. If we repeatedly perform the first two steps
of the measurement cycle, i.e.~prepare coherent states of the atomic
spins and performing the first measurement pulse with outcomes $A_1$
and $B_1$, we may deduce the statistical properties of the measurement
outcomes. Theoretically we expect from~(\ref{interact_light})
\begin{equation}
\label{eq:stat_projection}
  \mean{A_1} = \mean{B_1} = 0
  \quad\text{and}\quad
  \var(A_1) = \var(B_1) = \frac{1}{2} + \frac{\kappa^2}{2}. 
\end{equation}
The first term in the variances is the shot noise (SN) of light.  This
can be measured in absence of the interaction where $\kappa = 0$. The
quantum nature of the shot noise level is confirmed by checking the
linear scaling with photon number of the pulse, see
Eq.~(\ref{eq:shot_noise}). The second term arises from the projection
noise (PN) of atoms. Hence, we may calibrate $\kappa^2$ to be the
ratio $\kappa^2 = \mathrm{PN/SN}$ of atomic projection noise to shot
noise of light. Theoretically $\kappa^2$ has the linear scaling
$\kappa^2 = a J_x S_x T$ with the macroscopic spin size $J_x$ which
must be confirmed in the experiment.

Next we describe how to deduce the statistical properties of the state
created by the $\emph{entangling pulse}$. Based on the measurement
results $A_1$ and $B_1$ of this pulse we must predict the mean value
of the second measurement outcome. If $\kappa \rightarrow \infty$ we
ought to trust the first measurement completely since the initial
noise of $\X{Li}^{\mathrm{in}}$ is negligible, i.e.~$\mean{A_2} = A_1$
and $\mean{B_2} = B_1$. On the other hand, if $\kappa = 0$ we know
that atoms must still be in the vacuum state such that $\mean{A_2} =
\mean{B_2} = 0$. It is natural to take in general $\mean{A_2} = \alpha
A_1$ and $\mean{B_2} = \alpha B_1$. We need not know a theoretical
value for $\alpha$. The actual value can be deduced from the data. If
we repeat the measurement cycle $N$ times with outcomes $A_1^{(i)}$,
$B_1^{(i)}$, $A_2^{(i)}$, and $B_2^{(i)}$, the correct $\alpha$ is
found by minimizing the conditional variance
\begin{equation}
\label{eq:alpha}
\begin{split}
  \var(A_2|A_1) + \var(B_2|B_1) &= \\
  \min_{\alpha} \frac{1}{N-1}\sum_i^N &\left( 
    (A_2^{(i)} - \alpha A_1^{(i)})^2 + (B_2^{(i)} - \alpha B_1^{(i)})^2
    \right). 
\end{split}
\end{equation}
In order to deduce whether we fulfil the entanglement
criterion~(\ref{eq:ent_critetrion_xp}) we compare the above to our
expectation from~(\ref{interact_light}). For the verifying pulse we
get
\begin{equation}
  \begin{split}
   \mean{\left(\X{Li}^{\mathrm{out}}-\mean{\X{Li}^{\mathrm{out}}}\right)^2}
   &= \mean{\left(\X{Li}^{\mathrm{in,2nd}} + \kappa\left[
     \P{Ai}^{\mathrm{ent}} - \mean{\P{Ai}^{\mathrm{ent}}}\right]\right)^2}\\
   &= \frac{1}{2} + \kappa^2 \var(\P{Ai}^{\mathrm{ent}}),
  \end{split}
\end{equation}
where $\X{Li}^{\mathrm{in,2nd}}$ refers to the incoming light of the
\emph{verifying pulse} which has zero mean. $\P{Ai}^{\mathrm{ent}}$
refers to the atoms after being entangled.  We see that the practical
entanglement criterion becomes
\begin{equation}
\label{eq:easy_critertion}
  \begin{split}
      \var(A_2|A_1) + \var(B_2|B_1) 
      &= 1 + \kappa^2\left(
        \var(\P{A1}^{\mathrm{ent}})+\var(\P{A2}^{\mathrm{ent}})\right) \\
      < 1+\kappa^2 &= \var(A_1) + \var(B_1).
  \end{split}
\end{equation}
In plain English, we must predict the outcomes $A_2$ and $B_2$ with a
precision better than the statistical spreading of the outcomes $A_1$
and $B_1$ with the additional constraint that $A_1$ and $B_1$ are
outcomes of quantum noise limited measurements.
\subsection{Theoretical Entanglement Modeling}
\label{sec:theo_ent_model}
Above we described the experimental procedure for generating and
verifying the entangled states. Here we present a simple way to derive
what we expect for the mean values (i.e.~the $\alpha$-parameter) and
for the variances $\var(\P{Ai}^{\mathrm{ent}})$.

We calculate directly the expected conditional variance of $A_2$ based
on $A_1$:
\begin{equation}
  \begin{split}
    &\mean{\left(
    \X{L1}^{\mathrm{out,2nd}}-\alpha\X{L1}^{\mathrm{out,1st}}\right)^2} \\ 
   = &\mean{\left(\X{L1}^{\mathrm{in,2nd}}-\alpha\X{L1}^{\mathrm{in,1st}}
   + \kappa\left[\P{A1}^{\mathrm{in}}-\alpha\P{A1}^{\mathrm{ent}}\right]
    \right)^2} \\
   = &\frac{1}{2}(1+\alpha^2 + \kappa^2(1-\alpha)^2).
  \end{split}
\end{equation}
In the second step we assumed that the measurement is perfectly QND
and without any decoherence, i.e.~$\P{A1}^{\mathrm{ent}} =
\P{A1}^{\mathrm{in}}$. By taking the derivative with respect to
$\alpha$ we obtain the theoretical minimum
\begin{equation}
\begin{split}
\label{eq:var_theory}
  \var(A_2|A_1) + \var(B_2|B_1) &= 1 + \frac{\kappa^2}{1+\kappa^2} \\
  \Rightarrow  \var(\P{A1}^{\mathrm{ent}})+\var(\P{A2}^{\mathrm{ent}})
  &= \frac{1}{1+\kappa^2}
\end{split}
\end{equation}
obtained with the $\alpha$-parameter
\begin{equation}
\label{eq:alpha_theory}
  \alpha = \frac{\kappa^2}{1+\kappa^2}.
\end{equation}
It is interesting that in principle any value of $\kappa$ will lead to
creation of entanglement. The reason for this is our prior knowledge
to the entangling pulse. Here the atoms are in the coherent state
which is as well defined in terms of variances as possible for
separable states. We only need an ``infinitesimal'' extra knowledge
about the spin state to go into the entangled regime.

It is interesting to see what happens to the conjugate variables
$\X{Ai}$ in the entangling process. This is governed by
Eq.~(\ref{interact_atom}). We do not perform measurements of the light
operator $\P{Li}^{\mathrm{in}}$ so all we know is that both
$\X{Ai}^{\mathrm{in}}$ and $\P{Li}^{\mathrm{in}}$ are in the vacuum
state. Hence $\var(\X{Ai}^{\mathrm{ent}}) = (1+\kappa^2)/2$ and we
  preserve the minimum uncertainty relation
  $\var(\X{Ai}^{\mathrm{ent}})\var(\P{Ai}^{\mathrm{ent}}) = 1/4$.

\subsection{Entanglement Model With Decoherence}
Practically our spin states decohere between the light pulses and also
in the presence of the light. We model this decoherence naively by
putting the entire effect between the two pulses, i.e.~we assume there
is no decoherence in presence of the light but a larger decoherence
between the pulses. We may then perform an analysis in complete
analogy with the above with the only difference that
$\P{A1}^{\mathrm{ent}} =
\beta\P{A1}^{\mathrm{in}}+\sqrt{1-\beta^2}\vac{p}$ where $\vac{p}$ is
a vacuum operator admixed such that $\beta = 0$ corresponds to a
complete decay to the vacuum state and $\beta = 1$ corresponds to no
decoherence. Completing the analysis we find the theoretical
conditional variances
\begin{equation}
\begin{split}
\label{eq:var_theory_decoh}
  \var(A_2|A_1) + \var(B_2|B_1) &= 1 + 
    \kappa^2\frac{1+(1-\beta^2)\kappa^2}{1+\kappa^2} \\
  \Rightarrow  \var(\P{A1}^{\mathrm{ent}})+\var(\P{A2}^{\mathrm{ent}})
  &= \frac{1+(1-\beta^2)\kappa^2}{1+\kappa^2}
\end{split}
\end{equation}
obtained with $\alpha$-parameter
\begin{equation}
\label{eq:alpha_theory_decoh}
  \alpha = \frac{\beta\kappa^2}{1+\kappa^2}.
\end{equation}
In the limit $\beta \rightarrow 1$ these results agree
with~(\ref{eq:var_theory}) and~(\ref{eq:alpha_theory}). For $\beta
\rightarrow 0$ we have $\alpha \rightarrow 0$ (outcomes $A_1$ and
$B_1$ are useless) and the variance approaches that of the vacuum
state which is a separable state.

\section{Experimental Setup}
\label{sec:experimental-setup}
\begin{figure}[t]
\centerline{\includegraphics[width=\linewidth]{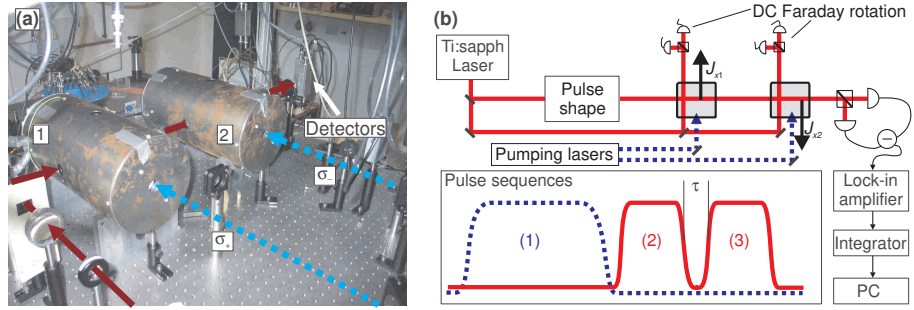}}
\caption{\small 
  {\bf(a)} A photographic view of the experimental setup. Atomic
  vapour cells are placed inside the cylindrical magnetic shields. The
  pumping beams are indicated with dashed arrows and the path of the
  entangling and verifying pulse is marked with the solid arrows. {\bf
    (b)} A schematic view of the setup. The pulses reach a detection
  system measuring $\S{y}(t)$. The photocurrent is sent to a lock-in
  amplifier which singles out the $\sin(\Omega t)$ and $\cos(\Omega
  t)$ parts. These are integrated and stored in a PC. The pulse
  sequence consists of (1) a 4ms pumping pulse, (2) a 2ms entangling
  pulse, a small delay $\tau = 0.25\textrm{ms}$, and (3) a 2ms
  verifying pulse. In addition to the pumping lasers and the
  entangling and verifying pulse, a laser beam is sent through each
  sample to measure the magnitude $J_x$ of the macroscopic spins by
  Faraday rotation measurements.}
 \label{fig:timingdet}
\end{figure}

In this section we describe the details of the experimental setup,
e.g.~laser settings, pulse lengths, detection systems, etc.  A picture
of the experimental setup is shown in Fig.~\ref{fig:timingdet}. In
part (a) we see two cylindrical magnetic shields which each contain a
paraffin coated vapour cell with cesium. The distance between the two
cells is 35cm.  In part (b) of the figure we show schematically the
timing of laser pulses and the detection system setup.

\subsection{Laser Settings and Pulse Timing}
\label{sec:meascycle}

In one measurement cycle the first step is to create the coherent spin
state of the atomic samples. To this end we have two diffraction
grating stabilized diode lasers, one at the 894nm D1 transition
$6S_{1/2}\rightarrow 6P_{1/2}$ and one at the 852nm D2 transition
$6S_{1/2}\rightarrow 6P_{3/2}$. We call these the optical pump and the
repump, respectively. Both are sent through the first gas sample along
the $x$-axis with $\sigma_+$ polarization, thus driving $\Delta m=+1$
transitions only.  In the second gas sample the polarization is
$\sigma_-$. The main pumping is done by the optical pump laser, which
drives the atoms towards the $F=4$, $m_F=4$ state (for
$\sigma_+$-polarization).  This state will be unaffected by the
optical pumping laser (a dark state) because of the absence of an
$F=5$ state in the $6P_{1/2}$ multiplet.  In the pumping process some
of the atoms will decay into the $F=3$ ground state, which is why we
need the repumping laser from this state to the $6P_{3/2}$ state to
return them to the pumping cycle.  Note that we can to some extent
control the number of atoms in the $F=4$ ground state with the power
of the repump laser. With this optical pumping scheme we can obtain
spin polarization above 99\% (measured by methods similar to
\cite{mors}).  The pumping pulses last for 4ms and are shaped by
Acousto Optical Modulators.

The atomic samples are now prepared in the CSS with anti-parallel
macroscopic orientation. Next the off-resonant entangling pulse shaped
by an Electro Optical Modulator of duration 2ms is sent through both
cells to the $\sh_y$ detection. The difference signal is fed into a
lock-in amplifier and the result is two numbers $A_1$ and $B_1$
corresponding to the integrated $\J{y1}'+\J{y2}'$ and
$\J{z1}'+\J{z2}'$ signals with an additional light contribution from
the incoming $\S{y}^{\mathrm{in}}(t)$. The entangling (and verifying)
pulses have power of $P = 4.5$mW and are blue detuned by 700MHz
compared to the $6S_{1/2}, F=4 \rightarrow 6P_{3/2}, F'=5$ transition.
They emerge from a Microlase Ti:sapphire laser which is pumped by an
8W Verdi laser.

After the entangling pulse is a short $\tau = 0.25$ms delay before the
verifying pulse is sent through the atoms. The entangling and
verifying pulses have exactly the same shape and duration. The
verifying pulse will again result in two numbers $A_2$ and $B_2$
stored in a PC. We now must predict the outcomes $A_2$ and $B_2$ based
on $A_1$ and $B_1$ as described in section~\ref{sec:ent_create_model}.
In order to calculate the statistics properly we repeat the
measurement cycle 10.000 times.

\subsection{Measuring the Macroscopic Spin}
In addition to the entanglement creation and verification we also
measure the macroscopic value $J_x$ of the spin samples by sending
linearly polarized light along the direction of optical pumping in
both samples. This light experiences polarization rotation
proportional to $J_x$. As already noted we have for small angles
$\theta = S_y/2S_x$. We also know $S_y = a S_x J_x$ (for small angles,
$S_y^{\mathrm{in}}$ is zero classically) when light is propagating
along the $x$-direction. Holding these together we find (this holds
for all angles)
\begin{equation}
\label{eq:theo_faraday_rot}
  \theta[\mathrm{rad}] = \frac{a J_x}{2} = 
    -\frac{\gamma\lambda^2 J_x}{16\pi A_{\mathrm{eff}}\Delta}.
\end{equation}
Instead of the beam cross section $A$ we need to use the
\emph{effective} cross section $A_{\mathrm{eff}} = 6.0\mathrm{cm}^2$
which is such that the volume of the vapour cell is $V =
A_{\mathrm{eff}}\cdot L$ where $L$ is the length traversed by the
laser beam (we call it ``effective'' since the vapour cell is not
exactly box like). We only measure polarization rotation from atoms
inside the beam cross section (which does not fill the whole volume)
and the equation must be scaled in order to count $J_x$ for all atoms
(we have $J_x = J_x^{\textrm{in beam}}\cdot A_{\mathrm{eff}}/A$).

The DC Faraday rotation angle $\theta$ is a practical handle on the
macroscopic spin size $J_x$ but we may in addition calculate a
theoretical level for the projection noise to shot noise ratio
$\kappa^2 = \mathrm{PN/SN}$. Theoretically $\kappa^2 = a^2 J_x S_x T$
where again it is a question which cross section $A$ to insert in $a$.
The entangling and verifying pulses do also not fill the entire vapour
cell volume. We assume that the measurement results are not depending
much in the real cross section for the following reason. If $A$ is
small the interaction with the atoms inside the beam is stronger, but
each atom will spend less time inside the beam leading to a
correspondingly shorter effective interaction time $T$ (atoms are at
room temperature and move in and out of the beam). If this simple
consideration has some validity we may assume that the light fills
the whole cell volume and the correct cross section to insert is the
effective area $A_{\mathrm{eff}}$. Now this can be hold up
against~(\ref{eq:theo_faraday_rot}). Inserting $\lambda = 852$nm,
$\gamma = 5$MHz and expressing the photon flux $\phi = 2S_x$ in terms
of probe power we may derive
\begin{equation}
  \label{eq:kappasqr_theory}
  \kappa^2_{\mathrm{theory}} = \frac{18.6\cdot P[\mathrm{mW}]
    \cdot T[\mathrm{ms}]\cdot\theta[\mathrm{deg}]}{\Delta[\mathrm{MHz}]}.
\end{equation}
We remember the theory is a bit crude but we will see in
Sec.~\ref{sec:experimental-results} that the model holds pretty well.

\section{Experimental Results}
\label{sec:experimental-results}
\begin{figure}[t]
\centerline{\includegraphics[width=0.7\textwidth]{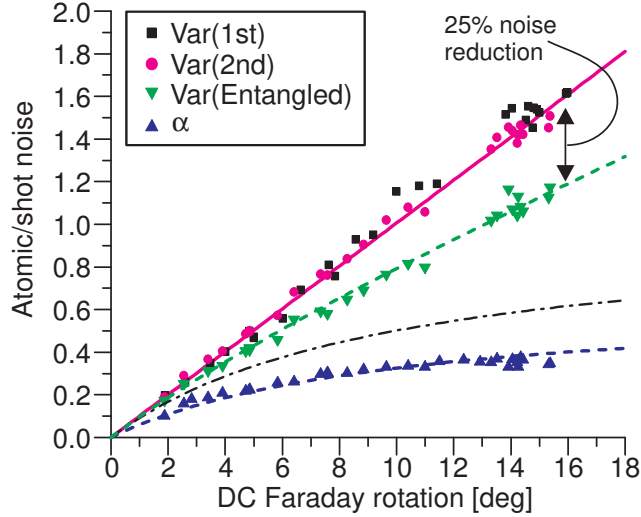}}
 \caption{\small 
   Atomic noise in units of shot noise as a function of atomic density
   (measured by DC Faraday rotation). Shot and electronics noise has
   been subtracted. Squares show the 1st pulse noise, circles the 2nd
   pulse noise. The linearity of these data is a finger print of the
   projection noise level which is then given by the solid line linear
   fit. Tip down triangles show the noise of the entangled states. Tip
   up triangles show the weight factor $\alpha$.  The two dashed
   curves trough triangles is the model described in
   Eqs.~(\ref{eq:var_theory_decoh}) and~(\ref{eq:alpha_theory_decoh})
   with $\beta = 0.65$. The dash-dotted curve is the theoretically
   best for the triangles ($\beta = 1$). Note that the states created
   by the first pulse measurements are really entangled states
   (according to the criterion~(\ref{eq:easy_critertion})) since the
   noise is clearly below the PN. We observe up to 25\% noise
   reduction. Note, entanglement is observed for low densities also
   with $\kappa^2 < 1$.}
\label{fig:alldataentanglement}
\end{figure}
The experimental data is shown in Fig.~\ref{fig:alldataentanglement},
in the following we carefully explain the details of this graph. 

\subsection{The Projection Noise Level}
The graph in Fig.~\ref{fig:alldataentanglement} has on the abscissa
the measured DC Faraday rotation angle which is proportional to $J_x$.
The squares are the variances $\var(A_1) + \var(B_1)$ for the
entangling pulses with shot noise (and electronics noise) subtracted.
Also, the results are normalized to shot noise. The circles are the
variances $\var(A_2) + \var(B_2)$ for the verifying pulses with the
same normalization. According to Eq.~(\ref{eq:stat_projection}) we
have plotted the experimental ratio $\kappa^2 = \mathrm{PN/SN}$ (the
unity term in~(\ref{eq:stat_projection}) is the subtracted shot
noise). The linearity with $J_x$ confirms that we measure projection
noise of atoms and not extra classical noise. We have roughly
$\kappa^2_{\mathrm{exp}} = 0.10\cdot \theta[\mathrm{deg}]$ which
should be compared to the prediction~(\ref{eq:kappasqr_theory})
$\kappa^2_{\mathrm{theory}} = 0.24\cdot\theta[\mathrm{deg}]$. The
discrepancy is a little more than a factor of two but this is
acceptable for our quite simple modeling. Note that the noise of the
verifying pulse is the same as that of the entangling pulse. This is
expected since we are performing a QND measurement. It is only when we
remember the information given to us by the measurement results $A_1$
and $B_1$ that we can tell more about the state created by the
entangling pulse.

\subsection{Conditional Variances and Entanglement}
The tip down triangles in Fig.~\ref{fig:alldataentanglement} is the
conditional variance $\var(A_2|A_1)+\var(B_2|B_1)$ normalized to shot
noise and with shot and electronics noise subtracted. According
to~(\ref{eq:easy_critertion}) we thus plot
$\kappa^2(\var(\P{A1}^{\mathrm{ent}})+\var(\P{A2}^{\mathrm{ent}}))$.
The fact that the points are lower than the straight line ($\kappa^2$)
is a direct indication that the entanglement
criterion~(\ref{eq:ent_critetrion_xp}) is fulfilled. For the higher
densities the reduction is 25\% but we note that entanglement is also
observed for smaller densities with $\kappa^2 < 1$. The latter was
impossible with the older methods applied in \cite{nature}. The
corresponding $\alpha$-parameters from the minimization
procedure~(\ref{eq:alpha}) are plotted in
Fig.~\ref{fig:alldataentanglement} with tip up triangles.

The expected entangled noise level in the ideal case is given
by~(\ref{eq:var_theory}). This is drawn as the dash-dotted curve
($\kappa^2$ times $1/(1+\kappa^2)$). We see the conditional variance
lies higher than this curve and hence the entanglement is worse than
expected.  According to~(\ref{eq:alpha_theory}) we also would expect
the $\alpha$-parameters to lie on the same dash-dotted curve in the
ideal case. It is clearly not the case, the experimental
$\alpha$-parameters are lower which indicates that the results $A_1$
and $B_1$ can not be trusted to as high a degree as expected.

Let us try to apply the simple decoherence model given by
Eqs.~(\ref{eq:var_theory_decoh}) and~(\ref{eq:alpha_theory_decoh}).
Taking the decoherence parameter $\beta = 0.65$ we get the dashed
lines in the figure. These match nicely the experimental data. We
conclude that the simple decoherence model has some truth in it and we
must accept that the entangled state created can only be verified to
be around ``65\% as good'' as expected in an ideal world.

\subsection{Physical Decoherence Processes}
Above we quantify the observed decoherence with the $\beta$-parameter.
Here we comment on the physical grounds for the decoherence. A well
known parameter for describing the decay of the \emph{transverse} spin
components $J_y$ and $J_z$ is the $T_2$-time defined by
\begin{equation}
\label{eq:T2decay}
  \frac{\partial J_i}{\partial t} = -\frac{J_i}{T_2}
\end{equation}
where $i=y,z$. We have studied the $T_2$-time extensively. A very good
method for this is to create (by applying an RF-magnetic pulse) a
displaced version of the coherent spin state with e.g.~$\mean{\J{y}'}
\ne 0$. This non-zero mean value can be detected by our standard
$\S{y}$-detection method and the decay following~(\ref{eq:T2decay})
may be observed.

Our experience tells us that power broadening by the laser pulses
combined with light assisted atomic collisions play the important role
in the decoherence processes. For high densities and high optical
powers we may find $T_2$ as low as 5ms. A fair guess for $\beta$ is an
exponential decay over a typical time scale of 2ms (the time between
the central parts of the entangling and verifying pulses). This yields
$\beta \approx \exp(-2\mathrm{ms}/5\mathrm{ms}) \approx 0.67$. This is
not far from the observed $\beta$ but we should say here that the 5ms
is a typical value not directly measured in the case of the data given
in Fig.~\ref{fig:alldataentanglement}. It is our experience that the
observed decoherence in the entanglement experiments is stronger than
that expected from the processes mentioned above. This is indeed true
for shorter pulses. We believe that the atomic motion in and out of
the laser beam combined with inhomogeneous light/atom coupling due to
the Doppler effect may also play a role, but we need further
investigation to confirm this completely.

\subsection{Conclusion of Experimental Results}
To conclude the experimental section we emphasize that we have
generated entangled states between distant atomic samples in the sense
that each vapour cell sits in its own magnetic shield. The two shields
can in principle be moved as far apart as is practical, our experiment
was performed with a distance of 35cm. In the future we hope to extend
this distance further. The noise reduction below the level set by
separable states was measured up to 25\%. This number is mainly
limited by power broadening and light assisted collisional relaxation
of the atomic spins. The decoherence is successfully modeled by a
single parameter $\beta$. A much more elaborate theory on entanglement
generation in presence of decoherence and losses exists \cite{myart}.
We should note, that the generated entangled states have random but
known (based on $A_1$ and $B_1$) mean values.  It is possible by
applying RF-magnetic fields to shift the entangled states to having
zero mean value while preserving the reduced variance. Experimental
demonstration of such will be considered elsewhere.
\section{Perspectives}
We will now present a brief overview of some of the interesting
applications in the field of quantum information of our reliable
source of distant atomic entanglement.
\subsubsection{Atomic Teleportation}
Quantum teleportation was first proposed in 1993 \cite{bennett} and
the year after for the special case of continuous variables
\cite{vaidman}.  Teleportation is extremely important since direct
transport of physical states is often hindered by exponential
decoherence. With quantum teleportation the information is cleanly
separated into a classical part, which can be transmitted over
arbitrary distances, and a quantum mechanical part, which only needs
to interact locally. 

\begin{figure}[t]
 \centerline{\includegraphics[width=\linewidth]{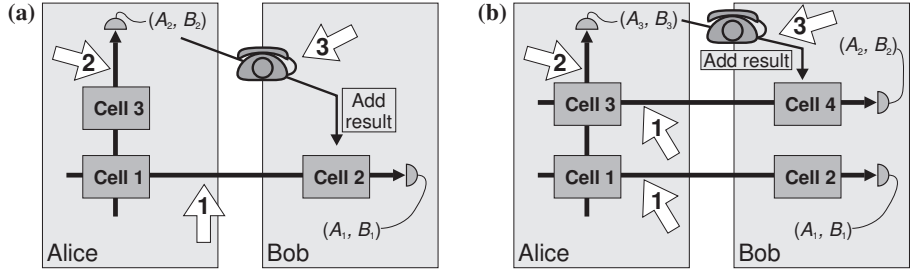}}
\caption{\small 
  {\bf (a)} Teleporting an unknown quantum state: first cells 1 and 2
  are entangled. Then Alice sends a light pulse through cell 1 and the
  unknown quantum state in cell 3. The measurement results are
  communicated classically to Bob who by applying a displacement to
  his system based on the results of the two measurement recreates the
  unknown quantum state in his cell. {\bf (b)} Entanglement swapping:
  the same procedure as in (a) but with two sets of entangled states
  initially. After a displacement based on the measurement results
  cells 2 and 4 are entangled even though they have never interacted
  directly.}
\label{fig:teleport}
\end{figure}

A proposal for spin state teleportation was given in \cite{Duan}.
Three atomic samples are needed as shown in
Fig.~\ref{fig:teleport}(a).  Adjacent samples are oriented oppositely
along the $x$-axis so that both collective measurements on cells 1 and
2 and on cells 1 and 3 will be regular entangling interactions as
discussed earlier. Cells 1 and 3 are located at Alice's site and cell
2 at Bob's site. The goal is now to teleport an unknown state in cell
3 onto cell 2. First cells 1 and 2 are entangled giving the
measurement results $(A_1,B_1)$. Next a pulse is sent through cells 1
and 3 and the measurement results $(A_2,B_2)$ are communicated
classically to Bob. He can now by applying a rotation to his system
based on the outcomes of the two measurements recreate the unknown
state in his cell. This of course only works with perfect fidelity in
the large interaction regime ($\ks\gg1$).

\subsubsection{Teleporting an Entangled State: Entanglement Swapping}
As described in \cite{royal} an entangled state can also be teleported
using macroscopic samples of atoms. In this case Alice and Bob each
have two samples (Fig.~\ref{fig:teleport}(b)). First each one of
Alice's samples are entangled with one of Bob's. Then a pulse of light
is sent through Alice's two samples making it an entangled state.
Alice now sends the result of a measurement on this last entangling
pulse to Bob. Using this and the results of the two primary entangling
pulses he can displace one of his states. This entangles his two
samples without ever bringing the two into direct contact. Had Alice
shifted one of her spin states prior to creating the entanglement
between her two cells the exact same protocol would allow Bob to
recover this shift in his cells. This allows for secret quantum
communication.

\subsubsection{Light to Atom Teleportation: Quantum Memory}
With an entangled source of atoms an unknown state of light can also
be teleported onto this \cite{kuzmich}. Assume we have two atomic
samples with $\Jh_{y1}+\Jh_{y2}=0$ and $\Jh_{z1}+\Jh_{z2}=0$, i.e.~a
perfect EPR-entangled state. The protocol is simpler without rotating
spins but also works with. As can be seen from Eq.~(\ref{eq:syout})
$\Jh_z$ is mapped onto $\sh_y$ when light propagating along the $z$
axis is sent through an atomic sample. If light is sent through the
first atomic sample and subsequently detected the measurement result
can be fed back into $\Jh_{z2}$ such that the original atomic
variables exactly cancel. $\sh_y$ has now been mapped perfectly onto
$\Jh_{z2}$.  Another effect of the light pulse is seen from
Eq.~(\ref{eq:jyout}).  $\sh_z$ is mapped onto $\Jh_{y1}$. If the
transverse spins of sample 1 are rotated $90^\circ$ and a new strong
light pulse ($\ks\gg1$) is sent through this sample
$\Jh_{y1}^{\mathrm{out}}$ is measured. This result can be fed back
onto $\Jh_{y2}$ in such a way that the two original $\Jh_y$'s cancel.
$\sh_z$ is now stored in $\Jh_{y2}$ and the teleportation is complete.
This process could also be reversed so that an unknown atomic state is
mapped onto a light pulse via two EPR-entangled light beams. This
shows that macroscopic samples of atoms offers a very feasible
protocol for complete quantum memory.

\section{Summary and Conclusion}
We have presented the first experimental realization of distant atomic
entanglement in the sense that the two atomic systems are placed in
separate environments, thus enabling entanglement between system
separated by arbitrary distances. Given the abundance of available
quantum information protocols for this type of continuous variable
atomic system the importance of this achievement is evident. Although
quantum teleportation of atomic states has been achieved recently
\cite{wineland,blatt} none of these approaches display directly
scalable distance between the unknown quantum state and the target
system. Our system must therefore be considered an important candidate
for achieving the long standing goal of high fidelity transfer of
atomic states over great distances upon which many of the proposed
technical applications of entanglement and teleportation critically
depend. The existence of several quantum information protocols for our
physical system is based on the simplicity of the interaction between
light and atoms as expressed by Eqs.~(\ref{interact_light}-d).

\subsection{Acknowledgments}
This work is supported by the Danish National Research Foundation and
the European Union through the grant QUICOV.

\bibliographystyle{phaip}
%\bibliography{distant}
\chapbblname{distant}%name of .bbl file (same as main file)
\chapbibliography{distant}%name of .bib file

\end{document}